\pdfoutput=1
\documentclass[aps, preprint, superscriptaddress, nofootinbib]{revtex4-1}
\usepackage{amsmath, amsfonts, amssymb, hyperref, color, graphicx, xspace}
\usepackage[T1]{fontenc}
\definecolor{nicered}{rgb}{.7,.1,.1}
\definecolor{nicegreen}{rgb}{.1,.5,.1}
\definecolor{darkblue}{rgb}{0,0,.5}
\hypersetup{colorlinks, citecolor=nicegreen, linkcolor=nicered, urlcolor=darkblue}

\begin{document}

\title{Exact one-loop false vacuum decay rate}

\author{Victor Guada}
\email{victor.guada@ijs.si}
\affiliation{Jo\v{z}ef Stefan Institute, Ljubljana, Slovenia}

\author{Miha Nemev\v{s}ek}
\email{miha.nemevsek@ijs.si}
\affiliation{Jo\v{z}ef Stefan Institute, Ljubljana, Slovenia}

\date{\today}

\begin{abstract}
\noindent 
We discuss an exact false vacuum decay rate at one loop for a real and complex
scalar field in a quartic-quartic potential with two tree-level minima.
The bounce solution is used to compute the functional determinant from both fluctuations.
We obtain the finite product of eigenvalues and remove translational zero modes.
The orbital modes are regularized with the zeta function and we end up with a
complete renormalized decay rate.
We derive simple expansions in the thin and thick wall limits and determine their validity.
\end{abstract}


% 03.70.+k 	Theory of quantized fields
% 11.10.-z	Field theory 
% 73.40.Gk	Tunneling

\maketitle

%
% Introduction
%
\section{Introduction} \label{secIntro}

% Introduction to tunneling in QFT, motivation
Tunneling phenomena are among the most fascinating physical processes. 
They initiate cosmological first order phase transitions, where an unstable ground state - a false 
vacuum (FV) - transforms into an energetically favorable one.
A bubble of true vacuum forms, expands quickly, collides with other bubbles and fills up the entire
universe.

% Semiclassical approximation and bounce solutions
Theoretical studies of such transitions were initiated by Langer~\cite{Langer:1967ax}
and applied to field theory by~\cite{Kobzarev:1974cp} and notably by 
Coleman~\cite{Coleman:1977py}.
The decay rate was shown to be
\begin{equation} \label{eqGamAB}
  \Gamma \propto A \, e^{-\mathcal S_0} \left(1 + \mathcal O\left(\hbar \right) \right) \, ,
\end{equation}
where $\mathcal S_0$ is the saddle point Euclidean action and $A$ is a dimensionful prefactor.

% High T, GWs, baryogenesis, B-fields
%
Understanding bubble nucleation is crucial for several reasons.
In the early universe, particles form a hot plasma, whose thermal effects can drive the 
transition~\cite{Linde:1980tt} and dynamically generate the observed dominance of matter over 
antimatter, e.g. in electroweak baryogenesis~\cite{Bochkarev:1990fx, Cohen:1990py, 
Turok:1990zg, Turok:1990in}.
Apart from creating baryons, colliding bubbles may produce observable gravitational wave 
signals~\cite{Witten:1984rs, Hogan:1986qda, Kosowsky:1992rz, Grojean:2006bp, 
Hindmarsh:2013xza, Cutting:2018tjt} and primordial magnetic fields~\cite{Vachaspati:1991nm, 
Sigl:1996dm, DeSimone:2011ek, Tevzadze:2012kk, Ellis:2019tjf}.
Current aLIGO~\cite{TheLIGOScientific:2014jea} and aVIRGO~\cite{TheVirgo:2014hva} observatories 
are operating at frequencies that are mostly insensitive to TeV scale first order phase transitions, but upcoming 
detectors, such as LISA~\cite{Audley:2017drz, Caprini:2019egz},
DECIGO~\cite{Kawamura:2011zz} and BBO~\cite{Crowder:2005nr, Corbin:2005ny},
will have the ability to test such scenarios.
Historically, vacuum stability played an important role in understanding the Higgs mass bounds 
of the Standard Model (SM) from stability~\cite{Weinberg:1976mh, Linde:1976ds} and 
longevity~\cite{Frampton:1976pb}; see the recent works~\cite{Isidori:2001bm, Andreassen:2017rzq} 
and references therein.

The semiclassical picture in~\eqref{eqGamAB} is analogous to the path integral in quantum 
mechanics~\cite{Feynman:1965, Kleinert:2004ev}, with the role of 
the classical trajectory replaced by the bounce.
The bounce is a nontrivial unstable~\cite{Derrick:1964ww} configuration that extremizes 
the action and interpolates between the two minima of the potential.
In~\cite{Coleman:1977py}, the bounce was found in the thin wall (TW) approximation,
valid when the two vacua are nearly degenerate.
It was also proven~\cite{Coleman:1977th} that the dominant contribution to the rate is
$O(4)$ symmetric, which simplifies the problem.
Finding a closed form solution is in general impossible because one is dealing with a 
stiff nonlinear second order differential equation.
Nonetheless, one can find exact solutions for specific potentials, including 
the Fubini-Lipatov instanton~\cite{Fubini:1976jm, Lipatov:1976ny} and its 
generalization~\cite{Loran:2006sf}, linear~\cite{Duncan:1992ai, Dutta:2012qt}, polygonal~\cite{Guada:2018jek}, 
logarithmic~\cite{Shen:1988si, FerrazdeCamargo:1982sk, Aravind:2014pva}, pure quartic~\cite{Lee:1985uv} and 
the quartic-quartic potential~\cite{Dutta:2011rc}.
The situation becomes more involved with multiple scalar fields, where
the bounce traverses a nontrivial path in field space.
Nevertheless, the problem is understood and a number of tools~\cite{Wainwright:2011kj, 
Masoumi:2016wot, Athron:2019nbd, Sato:2019wpo, Guada:2020xnz}
are available for a fast and stable evaluation of the action. 

Phenomenologically, piecewise potentials have been used to estimate the bounce solutions, where 
closed-form results are not readily available. 
The triangular potential~\cite{Duncan:1992ai} was employed in estimating metastable supersymmetric 
minima, e.g. in~\cite{Intriligator:2006dd, Carena:2012mw}, in the context of dark energy~\cite{Pastras:2011zr} and 
gravitational wave production~\cite{Jaeckel:2016jlh, Zhou:2020stj}.
The quartic-quartic potential~\cite{Dutta:2011rc, Dutta:2011ej} and other combinations were studied in 
inflationary settings~\cite{Dutta:2011rc}, where such glued bounce solutions are used~\cite{Kamada:2014ufa, 
Andrianov:2011fg} as well.
Other physically motivated examples~\cite{Gleiser:1990dt} of false vacuum decay include the non-perturbative effects in 2 and 3 
dimensions~\cite{Gleiser:1995tv, Gleiser:2007ts}.

% Fluctuations around the bounce, leading order corrections
The prefactor $A$ is a bit more challenging.
It involves the calculation of a functional determinant~\cite{Callan:1977pt}, related to the operator 
that describes quantum fluctuations around the bounce.
Not many closed form solutions exist, usually one estimates the dimensionful prefactor $A$ by 
the inverse radius of the bounce solution.
To get a more precise result, one can compute
the rate numerically at zero~\cite{Baacke:2003uw, Dunne:2005rt, Hur:2008yg} and
finite temperatures~\cite{Strumia:1998nf, Munster:1999hr}.
For multifields, the bounce action is still $O(4)$ symmetric~\cite{Blum:2016ipp} and recently 
progress was made in numerical calculations for gauge theories~\cite{Chigusa:2020pr}.

An analytic estimate for the prefactor in the TW limit was found in~\cite{Konoplich:1987yd}
(see also~\cite{Wipf:1985mr, Garbrecht:2015oea, Ai:2018guc}), while the issues with gauge and scale 
invariance of the unstable quartic in the SM were worked out recently~\cite{Isidori:2001bm, 
Endo:2017gal, Endo:2017tsz, Andreassen:2017rzq} and the Fubini-Lipatov was studied 
in~\cite{Garbrecht:2018rqx}.
The prefactor also cancels the renormalization scale-dependence of the Euclidean action and
stabilizes the bubble nucleation rate, see~\cite{Croon:2020cgk} for a recent work on 
the uncertainties regarding gravitational wave production.
However, to our knowledge, an exact solution for a potential with two separate tree level 
minima appears to be missing.
We fill this gap here by finding such a new result for the decay rate of a quartic-quartic 
potential and its complexified version.
We find a simple formula, where the energy scale of the FV factorizes and the rest of the prefactor
depends only on the ratios of vevs and quartic couplings between the false and true vacuum.
This setup can be considered as a benchmark for understanding the impact of finite 
one loop corrections, needed for a consistent evaluation of the total rate at one loop.
In particular, one can easily derive the behavior in the thin and thick wall limits, thereby providing 
generic expectations for the class of potentials, which are approximately 
scale-invariant around the two minima.

% Fluctuation determinants and some properties, need for renormalization, techniques to do it
There are a couple of subtleties that make the calculation of the functional determinant involved.
The rate depends on an infinite product of eigenvalues of the fluctuation operator around
the bounce, normalized to the FV ones.
It turns out that it is not necessary to find the complete eigensystem with fixed 
boundary conditions.
Instead, one can rely on the Gel'fand-Yaglom~\cite{Gelfand:1959nq} theorem and solve a 
related differential equation with Cauchy boundary conditions.
Evaluating it at the second boundary is equal to the product of eigenvalues.
This is a considerably simpler procedure, even when it cannot 
be performed in closed form, which is typically the case.

The resulting spectrum contains a single negative eigenvalue~\cite{Coleman:1987rm}
that describes the unstable direction of the expanding bubble.
In addition, any symmetry (translational, scale or internal global invariance) of the 
bounce is reflected in the number of zero eigenvalues~\cite{Kusenko:1995bw, Kusenko:1996bv}.
The quartic-quartic potential has no classical scale invariance, thus we only have to remove
the four translational zero modes, which is done by a perturbative deformation of the 
homogeneous solution.
This relates the dimensional prefactor to the energy scale in the theory and is proportional 
to the bounce action.

The final result is still infinite, as usual for quantities with a tree-level counterterm.
We regularize it by subtracting the divergent asymptotic terms, expanded in a consistent power counting scheme.
The asymptotic terms are added back in the UV, using the same power counting.
This can be done with the effective action and Feynman diagrams or via the zeta function.
We use the latter, where the zeta is defined via a contour integral~\cite{Kirsten:2007ev} and its 
asymptotic form is calculated perturbatively in powers of the fluctuation potential~\cite{Dunne:2006ct}.
We extend~\cite{Dunne:2006ct} to include the discontinuity of the quartic-quartic fluctuation
potential by going to higher orders in orbital eigenvalues to subtract all the infinities.

Finally, the UV terms need to be renormalized, either by computing the one loop counter-terms or 
by requiring the analyticity of the zeta function.
Both correspond to the same renormalization scheme, i.e. to dimensional regularization in the
$\overline{\text{MS}}$ and give the same answer for the single quartic case.

% Our work and how it is organized
We introduce the theoretical basis for the FV decay rate with quantum fluctuations in Sec.~\ref{sec:FVRate}.
In Sec.~\ref{sec:Bounce} we define the quartic-quartic potential, set up the notation and review the 
bounce field configuration and the Euclidean action.
Sec.~\ref{sec:FuncDet} deals with the functional determinants: the general formalism,
exact expressions for the product of eigenvalues, removal of zeroes and the finite sum.
In Sec.~\ref{sec:ZetaContour} we review the zeta function formalism via the contour integral, then 
calculate the expansion around the FV in Sec.~\ref{sec:ZetaRenorm} and get the finite and renormalized terms.
Final result and thin/thick wall expansions are summarized in Sec.~\ref{sec:Summary}, with
the complexified version in Sec.~\ref{subsec:CplxQ}.
The outlook for further developments is discussed in Sec.~\ref{sec:Outlook} and technical details 
are left to the appendices~\ref{sec:Approximations} and~\ref{sec:DetailsAsymp}.

%
% False vacuum decay rate
%
\section{False vacuum decay rate} \label{sec:FVRate}
The false vacuum decay rate was derived in~\cite{Callan:1977pt} (see also~\cite{Hammer:1978xu}) 
and explained in more detail in Coleman's lectures~\cite{Coleman:1978ae, Coleman:1985} and 
classic textbooks~\cite{Weinberg:2012pjx, Weinberg:1996kr}.
A recent rederivation~\cite{Andreassen:2016cff, Andreassen:2016cvx} used a more direct approach 
via the path integral formulation.
The decay rate per space-time volume can be written as
\begin{align} \label{eqGammaFV}
  \frac{\Gamma}{\mathcal V} &= 
  \frac{\text{Im} \int \mathcal D \varphi \, e^{-S[\overline \varphi]}}{
  		        \int \mathcal D \varphi \, e^{-S[\varphi_{\text{FV}}]}} = 
  \left(\frac{\mathcal S_0}{2 \pi} \right)^2 e^{- \mathcal{S}_0} \,
  \text{Im} \sqrt{ \frac{\text{det} \mathcal O_{\text{FV}}}{\text{det'} \mathcal O}}
  \left(1 + \mathcal O\left(\hbar \right) \right) \, ,
\end{align}
were $\int \mathcal D \varphi$ is the path integral over scalar field fluctuations and 
$S[\varphi]$ is the action functional in Euclidean space-time. 
The numerator is the path integral for real scalar fluctuations around the bounce field configuration, 
with an imaginary component, while the denominator is the FV normalization.

The Euclidean action $S[\varphi]$ is expanded around the bounce $\varphi = \overline \varphi + \psi$ to
second order 
\begin{align} \label{eqSEuclid}
  S[\varphi] &\simeq S[\overline \varphi] + \frac{1}{2} \psi \mathcal O \psi + \ldots ,
  &
  \mathcal O &= \frac{\delta^2 S}{\delta \varphi^2}[\overline \varphi] \, .
\end{align}
The first derivative is zero because $\overline \varphi$ extremizes the action.
The operator $\mathcal O$ is the fluctuation operator, defined as the second derivative of the 
action, evaluated on the bounce, while $\mathcal O_\text{FV}$ is given in the FV.

Expanding $\psi$ in a set of eigenfunctions $\psi_{\bf n}$ of the fluctuation operator 
$\mathcal O \psi_{\bf n} = \gamma_{\bf n} \psi_{\bf n}$, we perform the Gaussian integral in~\eqref{eqGammaFV} 
and end up with the ratio of functional determinants~\cite{Callan:1977pt}.
The final step is to remove the translational zero modes by integrating over the collective
coordinates~\cite{Gervais:1974dc, Gervais:1975pa, Callan:1975yy, Jevicki:1976kd,
Andreassen:2017rzq}, which produces the space-time volume factor $\mathcal V$ on the left hand 
side of~\eqref{eqGammaFV} and the $\sqrt{\mathcal S_0/(2 \pi)}$ for every dimension of space-time.
We thus end up with~\eqref{eqGammaFV}, where the prime in $\text{det'}$ corresponds to 
the four removed eigenvalues, each with dimension of mass$^2$, which gives the correct
dimension of the decay rate.

%
% Bounce
%
\section{The Bounce solution} \label{sec:Bounce}
The leading semiclassical approximation, i.e. the $\mathcal S_0$ coefficient in~\eqref{eqGamAB} 
and~\eqref{eqGammaFV}, is given by the bounce field configuration $\overline \varphi(\rho)$. 
The manifest dependence on $\rho^2 = t^2 + \sum x_i^2$ takes care of Euclidean spherical 
symmetry~\cite{Coleman:1977th}. 
Extremizing the action $S[\varphi]$ leads to the bounce equation
\begin{equation} \label{eqBounceEq}
  \ddot \varphi + \frac{3}{\rho} \dot \varphi = V^\prime(\varphi) \, ,
\end{equation}
where $V' = \text{d} V/\text{d} \varphi$ and with the boundary conditions $\dot {\overline \varphi}(0,\infty) = 0$, where the 
bounce interpolates between the true $\overline \varphi(0) = \varphi_0 \simeq \varphi_{\text{TV}}$ and the false vacuum 
$\overline \varphi(\infty) = \varphi_{\text{FV}}$.

Let us consider the exact bounce solution for the quartic-quartic potential
\begin{align} \label{eqDefV}
  V = \frac{1}{4} \left( \lambda_2 v_2^4 - \lambda_1 v_1^4 + 
  \lambda_1 \left( \varphi + v_1 \right)^4 \right) H{(-\varphi)} +
  \frac{\lambda_2}{4} \left( \varphi - v_2 \right)^4 H{(\varphi)} \, .
\end{align}
where $H$ is the step function.
The two segments are joined at $\varphi = 0$ into a continuous $V$ with
the minima located at $-v_1$ and $v_2$.
We assume that $v_2 > 0$ is the FV with $V(v_2) = 0$ 
to simplify the bounce calculation.
Furthermore, for $v_1 > 0$ to be the true vacuum (TV), we need $\lambda_{1,2} > 0$ and 
require $\lambda_1 v_1^4 > \lambda_2 v_2^4$ such that $V(v_1) < 0$.

The derivative of the potential that enters in~\eqref{eqBounceEq} is 
\begin{align} \label{eqdV}
  V^\prime &= \lambda_1 \left(\varphi + v_1 \right)^3 H{(-\varphi)} + 
  \lambda_2 \left(\varphi - v_2 \right)^3 H{(\varphi)} \, .
\end{align}
The Dirac deltas, coming from the derivative of the step function in~\eqref{eqDefV}, vanish due to 
the continuity of $V$ at $\varphi = 0$.
Such potential was studied in~\cite{Dutta:2011rc} and admits an exact solution consisting of two 
pieces, glued together at $\rho = R_T$
\begin{align} \label{eqbounce}
  \overline{\varphi} = \sum_{s=1}^2
  \left( (-1)^s v_s + \sqrt{\frac{8}{\lambda_s}} \frac{R_s}{R_s^2 - \rho^2}\right) H \left((-1)^s(  \rho-R_T) \right) \, .
\end{align}

For later convenience, we define two dimensionless quantities $x$ and $y$ from
$v_1 = x \, v_2 = x \, v,$ and $\lambda_1 = y \, \lambda_2 = y \, \lambda$.
Demanding the potential to be convex\footnote{One can also consider negative $\lambda$ and reproduce the SM 
instability, as we will see below.},
implies $x > 0, y > 0$ and $x^4 y > 1$. 
Near the equality $x^4 y \simeq 1$, we approach the TW limit, where the minima are degenerate and the rate vanishes. 
The bounce parameters $R_{1,2,T}$ are obtained by matching the solution to $\varphi = 0$,
and requiring $\overline{\varphi}$ to be continuous and differentiable at $\rho = R_T$.
The resulting Euclidean radii are
\begin{align} \label{eqR12T}
  R_{1,2,T} &= \frac{2}{v} \sqrt{\frac{2}{\lambda}} \, \frac{1 + x}{x^4 y - 1} 
  \left \{ x^2 \sqrt y, 1, \sqrt{\frac{x + x^4 y}{1 + x}} \right \}.
\end{align}
Their size is set by $1/v$, as expected on dimensional grounds since $v$ is the relevant mass scale.  
Moreover, the radii are positive and diverge in the TW limit $x^4 y \to 1^{+}$, where the tunneling rate goes to zero.
The resulting bounce action is 
\begin{align}\label{eq:Action}
  \mathcal S_0 &= 2 \pi^2 \int_0^\infty \text{d} \rho \, \rho^3 \left(
    \frac{1}{2} \dot{\overline \varphi}^2 + V\left(\overline \varphi \right) \right)
  \\
  &=\left( \frac{8 \pi ^2}{3 \lambda} \right) 
  \frac{1 + y + x^3 y \left(4 + x y \left(-3 + 6 x^2 + (3 + 4 x) x^4 y \right) \right)}{y (x^4 y - 1)^3} \,.
\end{align}
The factor of $8 \pi ^2/(3 \lambda)$ is the well-known single quartic result, which gets multiplied
by a function that diverges when $x^4 y \to 1$ and thus $\Gamma \propto e^{-\mathcal S_0} \to 0$.
With the bounce at hand, we can proceed to make sense of the quantum fluctuations.

%
% Fluctuations
%
\section{Functional determinants} \label{sec:FuncDet}
As discussed in Sec.~\ref{sec:FVRate}, we are interested in calculating the spectra of eigenvalues of the
fluctuation operator $\mathcal O$, appearing in~\eqref{eqSEuclid}.
To this end, we employ the radial decomposition in four dimensions and get the product
of eigenvalues for a fixed orbital momentum mode $l$, by use of the Gel'fand-Yaglom 
theorem~\cite{Gelfand:1959nq}.

%
% Radial mode separation and exact product of eigenvalues
%
\subsection{Radial mode separation and exact product of eigenvalues} \label{sub:Radial}
We would like to find the product of eigenvalues $\gamma_{\bf n}$, associated to $\mathcal O$
\begin{align}\label{eq:eigenvalueEquation}
  \mathcal O &= - \Box + V'' \left( \overline \varphi \right), 
  &
  \mathcal O \psi_{\bf n} &= \gamma_{\bf n} \psi_{\bf n} \, ,
\end{align}
where $\Box$ is the Laplace operator in flat 4D Euclidean space-time. 
Here, ${\bf n}$ is a collective index for the relevant quantum numbers that come about when 
the boundary conditions $\psi_{\bf n}(0) = \psi_{\bf n}(\infty) = 0$ are imposed. 
The fluctuation potential follows from~\eqref{eqdV}
\begin{align} \label{eqDefVpp}
    V'' &= 3 \left( \lambda_1 \left(\varphi + v_1 \right)^2 H(-\varphi) + 
    \lambda_2 \left(\varphi - v_2 \right)^2 H(\varphi) \right) - 
    \left( \lambda_1 v_1^3 + \lambda_2 v_2^3 \right) \delta(\varphi) \, ,
\end{align}
and contains a delta function due to the discontinuity of $V^\prime$ at the origin.
The $V^{\prime \prime}(\overline \varphi(\rho))$ is 4D symmetric, therefore we can separate the 
radial and orbital part of $\psi_{\bf n}$, where the latter is described by hyperspherical harmonics.
These are eigenfunctions of the total orbital momentum operator with orbital quantum numbers 
$l = 0, \ldots, \infty$ that are $(l + 1)^2$-fold degenerate~\cite{Kleinert:2004ev}.
According to the Gel'fand-Yaglom theorem~\cite{Gelfand:1959nq}, we have to find the zero eigenmode of 
the fluctuation operator 
\begin{align} \label{eqDefOl}
   \mathcal O_l \psi_l = -\ddot \psi_l - \frac{3}{\rho} \dot \psi_l + \frac{l(l+2)}{\rho^2} \psi_l + 
   V''\left( \overline \varphi \right) \psi_l &= 0 \, ,
\end{align}
and evaluate $\psi_l$ at the boundary when $\rho \to \infty$.
This gives the log of the ratio of determinants
\begin{align} \label{eqLogDetO}
  \ln \left(  \frac{ \det \mathcal O}{\det  \mathcal O^{\text{FV}}} \right) &= 
  \sum_{l=0}^\infty \left( l + 1 \right)^2 \ln \mathcal R_l\left( \infty \right) ,
  &
  \mathcal R_l & \equiv \frac{\psi_l}{\psi^{\text{FV}}_l} \, .
\end{align}

Let us see how the fluctuations behave.
In the FV, we have $V''_\text{FV} = 0$ and the solution of~\eqref{eqDefOl} is $\psi^\text{FV}_l = \rho^l$.
We dropped the part that diverges at $\rho = 0$ and assigned the arbitrary multiplication constant to 1.
The general solution of~\eqref{eqDefOl}, when the fluctuation potential is evaluated around the bounce, 
is instead given by 
\begin{align} \label{eqPsil}
\begin{split}
  \psi_{l s} & = A_{l s} \, \frac{\rho^l \, R_s^4}{\left(R_s^2 - \rho^2 \right)^2} \left( 1 - 2 \left( \frac{l - 1}{l + 2} \right) 
  \frac{\rho^2}{R_s^2} + \frac{l(l-1)}{(l+2)(l+3)} \frac{\rho^4}{R_s^4} \right)
  \\
  &
  + B_{l s} \, \frac{R_s^{l+4}}{(R_s^2 - \rho^2)^2} \frac{R_s^{l + 2}}{\rho^{l+2}} \left( 1 - 2 \left(\frac{l + 3}{l} \right) 
  \frac{\rho^2}{R_s^2} + \frac{(l+2)(l+3)}{l(l-1)} \frac{\rho^4}{R_s^4} \right) \, .
\end{split}
\end{align}
where $s = 1, 2$ denotes the two segments of the quartic-quartic potential.

On the first segment with $s=1$, regularity of $\psi_{l s}$ at $\rho = 0$ requires $B_{l1}$ = 0, and we choose
the normalization $A_{l1} = 1$, such that we normalize to the FV at $\rho = 0$.
This part reduces to the unstable single potential of the SM~\cite{Isidori:2001bm, 
Andreassen:2017rzq}, where we can easily read off the ratio $\mathcal R_l(\infty) = 
\lim_{\rho \to \infty} \psi_{l1}/\rho^l$ from the only term remaining in~\eqref{eqPsil} at high $\rho$
\begin{equation} \label{eqRlsingleq}
  \lambda \varphi^4 : \quad \mathcal R_l(\infty) 
  = \frac{\psi_{l1} \left(\infty \right)}{\psi^\text{FV}_l \left(\infty \right)} 
  = \frac{l(l-1)}{(l+2)(l+3)} \,.
\end{equation}

On general grounds~\cite{Coleman:1987rm}, we expect the $l=0$ mode to be negative, corresponding to 
the expanding bubble.
On the other hand, the four $l=1$ eigenvalues should vanish because of the translational invariance of the 
center of the bubble (or the bounce solution, which depends only on $\rho$).
The $\mathcal R_l(\infty)$ in~\eqref{eqRlsingleq} indeed contains a zero mode at $l=1$, but also has an 
additional zero at $l=0$, due to the classical scale invariance~\cite{Isidori:2001bm, Andreassen:2017rzq}.

Let us move on to the second segment and glue the two solutions.
The fluctuation potential contains a Dirac delta, therefore the derivative of $\psi_l$ changes 
discontinuously\footnote{Integrating~\eqref{eqDefOl} around $R_T$, we have
$\int_{R_T - \epsilon}^{R_T + \epsilon} \text{d} \rho \mathcal O_l \psi_l = 0 \xrightarrow{ \epsilon \to 0} 
\dot \psi_l (R_T + \epsilon) -\dot \psi_l (R_T -\epsilon) = -\left(\lambda_1 v_1^3 + 
\lambda_2 v_2^3 \right) \int_{R_T - \epsilon}^{R_T+\epsilon} \text{d} \rho \delta \left( \bar\varphi\left( \rho \right)\right) \psi_l$. }.
The appropriate boundary conditions to join $\psi_{l1,l2}$ at $\rho = R_T$ are given by
\begin{align} \label{eqMatchRl}
  \psi_{l1} &= \psi_{l2} \, , 
  &
  \dot \psi_{l1} &= \dot \psi_{l2} + \mu_V \psi_{l1} \, ,
  &
  \mu_V  &= \frac{\lambda_1 v_1^3 + \lambda_2 v_2^3}{\dot{ \overline \varphi}(R_T)} \, .
\end{align}
These fix the remaining parameters $A_{2l}, B_{2l}$ that ultimately determine the behavior of $\mathcal R_l$ 
as $\rho \to \infty$. 
We arrived to our main result for the fluctuation determinant  
\begin{align} \label{eqRlinf}
  \mathcal R_l(\infty) &= A_{l2} \frac{l(l-1)}{(l+2)(l+3)} = \frac{(l-1) (l^3 + c_2 l^2  + c_1 l+ c_0)}{(l+1) (l+2)^2 (l+3)} \, ,
\end{align}
with the three coefficients $c_i$ that depend only on dimensionless ratios $x$ and $y$:
\begin{align} \label{eqc0}
  c_0 &= \frac{12 (1+x)^2 x ^4 y (1 + x^3 y)^2}{ (x^4 y - 1)^3} \, ,
  \\ \label{eqc1}
  c_1 &= \frac{2 x \left( 1 + \left(1 + 2 x \right) x^2 y \right) \left( 2 + 3 x + (3 + 4 x) x^3 y\right)}{ (x^4 y - 1)^2} \, , 
  \\ \label{eqc2}
  c_2 &= \frac{1 + 4 x + (4 + 7 x) x^3 y}{x^4 y - 1} \, .
\end{align}
All $c_i$ are real and positive because $x^4 y > 1$, which follows from the construction of the potential.
Similarly to the radii $R_{1,2,T}$, the $c_i$ diverge in the TW limit.

The zero eigenvalue of the scale invariant single quartic in~\eqref{eqRlsingleq} at $l=0$ is now gone,
it got absorbed by the $A_{l2} \propto l$ in~\eqref{eqRlinf}.
This happens because the quartic-quartic contains mass scales $v_{1,2}$ that 
break scale invariance, thereby the $l=0$ mode in~\eqref{eqRlinf} becomes negative
\begin{align}\label{eq:Rl0infity}
\mathcal{R}_0(\infty) &= - \frac{c_0}{12} < 0 \,,
\end{align}
as required from the instability of the bounce solution.

It follows from~\eqref{eqRlinf} that $\mathcal R_l(\infty) \xrightarrow{l \gg 1} 1$ and the sum over $l$
in~\eqref{eqLogDetO} diverges quadratically in the UV - after all, we are computing a one loop quantity
with a tree level counterterm.
In Sec.~\ref{sec:ZetaRenorm} we will regularize the sum by subtracting the terms divergent in $l$
and calculate the finite part.
Before that, let us deal with the removal of the translational zero eigenvalues of the $l=1$ modes.

%
%  Removing the zero modes
%
\subsection{Removing the zero modes} \label{sub:RemoveZero}
As discussed in Sec.~\ref{sec:FVRate}, the prefactor is proportional to the reduced determinant,
where the four translational zero eigenvalues are subtracted.
The reduced contribution from the $l=1$ modes is defined as
\begin{equation} \label{eqR1p}
   \mathcal O_l \psi_l = \gamma_n \psi_l \implies
  \mathcal R^\prime_1(\infty) = \frac{\prod_{n = 2}^\infty \gamma_n}{\prod_{n = 1}^\infty \gamma^{\text{FV}}_n} \, .
\end{equation}
Omitting the zero modes is a straightforward procedure when $\gamma_n$ are known for the principal 
quantum numbers $n$.
However, with the Gel'fand-Yaglom approach, the eigenvalues are regrouped in terms of orbital $l$ modes.
Thus the zero from translations has to be removed carefully because it multiplies all 
the other eigenvalues with $l=1$.
This can be done perturbatively~\cite{Jevicki:1976kd, Endo:2017gal, Endo:2017tsz, Andreassen:2017rzq} by off-setting the 
fluctuation potential with a small dimensionful parameter $\mu^2_\varepsilon$ and finding the corresponding 
eigenfunctions of
\begin{equation} \label{eqOeps}
  \left( \mathcal O_1 + \mu^2_\varepsilon \right) \psi_1^\varepsilon = 0 \, .
\end{equation}
Instead of approaching zero, the ratio of determinants is then given by
\begin{equation}
  \mathcal{R}^\varepsilon_1(\infty) = \frac{\psi_1^\varepsilon(\infty)}{\psi^\text{FV}_1(\infty)} 
  \simeq \frac{ \left(\mu^2_\varepsilon + \gamma_1 \right) 
  \prod_{n=2}^\infty  \gamma_n}{\prod_{n=1}^\infty \gamma^{\text{FV}}_n}
  = \mu^2_\varepsilon \mathcal R^\prime_1(\infty) \, ,
\end{equation}
because the $\mu^2_\varepsilon$ shift does not affect $\gamma_{n>1}$ and $\gamma^{\text{FV}}_n$. 
In other words, we need to compute
\begin{equation} \label{eqR1pDef}
  \mathcal{R}'_1(\infty) = \lim_{\mu^2_\varepsilon \to 0} \frac{1}{\mu^2_\varepsilon} \mathcal{R}^\varepsilon_1(\infty) \, .
\end{equation}
The eigenfunctions $\psi_1^\varepsilon$ are infinitesimally close to $\psi_1$ and we can perform a perturbative 
expansion $\psi_1^\varepsilon \simeq \psi_1 + \mu^2_\varepsilon \, \delta \psi_1$, which enters 
in~\eqref{eqOeps}, such that
\begin{align} \label{eqOepsPsieps}
  \left(\mathcal O_1 + \mu_\varepsilon^2 \right) \left( \psi_1 + \mu_\varepsilon^2 \delta \psi_1 \right) \simeq  
  \mathcal O_1 \psi_1 + \mu_\varepsilon^2 \left( \psi_1 + \mathcal O_1 \delta \psi_1 \right) = 0 \,.
\end{align}
The general solution $\psi_{ls}$ in~\eqref{eqPsil} is singular for $l=1$, so we rederive it 
\begin{align} \label{eqPsi1}
  \psi_{1 s} =  \frac{R_s^4 \rho}{ \left( R_s^2 - \rho^2\right )^2} \left( A_{1s} + B_{1s} \left(
  \frac{\rho^4}{R_s^4} - 8 \frac{\rho^2}{R_s^2} + 24 \log \rho + 8 \frac{R_s^2}{\rho^2} - \frac{R_s^4}{\rho^4} \right) \right)\, .
\end{align}
On the first segment with $s=1$, the $\psi_{11}$ needs to be regular at $\rho = 0$ and normalized to the 
FV, therefore $A_{11} = 1$ and $B_{11} = 0$. Matching to the second segment at $R_T$ gives 
$A_{12} = x^6 y^2$ and $B_{12} = 0$.
The value at infinity is then given by $\psi_{12}(\infty) \propto B_{12} = 0$, as it should be
since we are looking at the zero eigenvalue and $\mathcal R_1(\infty) \propto \psi_{12}(\infty) = 0$.

Now that we have the $l=1$ fluctuation, let us move on to perturbations $\delta \psi_{1s}$,
given by the nonhomogeneous equation $\mathcal O_1 \delta \psi_1 = -\psi_1$ that comes 
from~\eqref{eqOepsPsieps} and get
\begin{equation}\label{eqdeltaPsi1}
\begin{split} 
  \delta\psi_{1 s} = \frac{3 R_s^6 \rho}{4 \left( R_s^2 - \rho^2\right )^2} \biggl( & 
  \delta A_{1s} + \frac{\delta B_{1s}}{18} \left(
  \frac{\rho^4}{R_s^4} - 8 \frac{\rho^2}{R_s^2} + 24 \log \rho + 8 \frac{R_s^2}{\rho^2} - \frac{R_s^4}{\rho^4} \right) -
  \\
  & \frac{A_{1s}}{18} \left( 6 \frac{\rho^2}{R_s^2} - 18 - 24 \log \rho - \frac{R_s^2}{\rho^2} + \frac{R_s^4}{\rho^4} \right)
  \biggr) \, .
\end{split}
\end{equation}
The boundary conditions $\delta\psi_{11}(0) = \dot {\delta\psi}_{11}(0) = 0$ fix $\delta A_{11} = \delta B_{11} = -1$ on 
the first segment\footnote{The single quartic case $\lim_{\rho \to \infty} \delta\psi_{11}/\rho = -R^2/24$ 
becomes consistent with the SM~\cite{Isidori:2001bm, Andreassen:2017rzq} after flipping the sign of 
$V^{\prime \prime}$, because we assumed $\lambda > 0$.}
 and the same matching conditions required for $\psi_l$ in~\eqref{eqMatchRl}, apply to $\delta \psi_l$.
These determine the remaining $\delta A_{12}$ and $\delta B_{12} = 3 \lambda/(8 \pi ^2) \mathcal S_0 x^6 y^2$.

In fact, it is $\delta B_{12}$ that gives the reduced determinant after plugging the 
expansion in~\eqref{eqR1pDef}
\begin{equation} \label{eqR1pFin}
  \mathcal R^\prime_1(\infty) = \lim_{\mu^2_\varepsilon \to 0} \frac{1}{\mu^2_\varepsilon} 
  \frac{\psi_1 + \mu^2_\varepsilon \delta \psi_1}{\psi_{\text{FV} 1}} \Bigr|_{\rho = \infty} 
  =  \frac{\delta \psi_1}{\psi_{\text{FV} 1}} \Bigr|_{\rho = \infty} 
  = \frac{R_2^2}{24} \delta B_{12} 
  = \frac{R_2^2}{24} \left( \frac{3 \lambda}{8 \pi ^2}\right) \mathcal S_0 x^6 y^2 \, ,
\end{equation}
where we used the fact that $\psi_1(\infty) = 0$ and $R_2$ was calculated in~\eqref{eqR12T}.
Note that the $\mathcal R^\prime_1$ is proportional to $\mathcal S_0$, which cancels with the prefactor 
in~\eqref{eqGammaFV}.
The reduced determinant has the correct dimension of mass$^{-2}$, set by the dimensional
$R_2$, which in turn is proportional to $1/v$, the energy scale of the model.
The dimensionless $\delta B_{12}$ serves as the numerical prefactor
that diverges in the TW limit and gives an additional suppression to the rate.
With the $l=1$ zero removed, we can proceed to the finite part.

%
% Regularization of the Sum
%

\subsection{Finite sum}\label{sub:FiniteSum}
With $\mathcal R_l$ in~\eqref{eqLogDetO} at hand, the finite part can be computed in some generality.
Let us consider a generic form of $\mathcal R_l$, given by a ratio of polynomials of order $n$
\begin{align} \label{eqRlab}
  \mathcal R_l(\infty) = \prod_{i=1}^{n}\frac{l + 1 - a_i}{l + 1 - b_i} \, ,
\end{align}
which covers the two cases in~\eqref{eqRlsingleq} and~\eqref{eqRlinf}.
The number of roots and poles must be the same, a consequence of the normalization to
the FV in~\eqref{eqLogDetO}.
To get the finite part of~\eqref{eqLogDetO}, we first find the asymptotic behavior of $\mathcal R_l$  
by expanding the log of the determinant for large $l$.
The degeneracy factor goes as $l^2$, therefore the $\ln \mathcal R_l$ has to be expanded up to $1/l^3$ 
to account for the quadratic, linear and log divergencies.

It turns out that the asymptotics of the zeta function, used for renormalization, will be given in powers of $\nu = l+1$, 
therefore it is convenient to define $\mathcal R_l^a$ by expanding~\eqref{eqLogDetO} in $1/\nu$ up to 
$\mathcal O \left( \nu^{-3} \right)$.
This is subtracted from~\eqref{eqLogDetO} and we get
\begin{align} \label{eq:zetafiniteRl}
  \Sigma_f &= \sum_{\nu = 1}^\infty \nu^2 \left( \ln \mathcal R_l(\infty) - \ln \mathcal R_l^a(\infty) \right),
\end{align}  
which is convergent and can be computed\footnote{Technically, we do the sum over $\mathcal R_l(\infty)$
from $\nu = 3$ up to a large finite regulator to skip the $l=0,1$ modes, which are then added by hand. 
The sum over $\mathcal R_l^a(\infty)$ starts from $\nu = 1$ as in the renormalization procedure.
After the summation, the regulator disappears and can be taken to infinity.} 
in a closed form
\begin{align}
  \begin{split}\label{eqSigmafinite}
  \Sigma_f &=  \sum_{i=1}^n \biggl( 
   \frac{a_i^3}{3} \gamma_E - \frac{a_i}{12} \left( 1+ 3 a_i - 6 a_i^{2}  \right) 
   - \zeta_R' \left(-2, 3 - a_i \right) - 2 a_i \zeta_R' \left(-1, 3 - a_i \right) 
  \\
  & - a_i^2 \zeta_R' \left(0, 3 - a_i \right) - \left( a \to b \right) \biggr) +  
  \ln \mathcal R_0(\infty) + 4 \ln \mathcal R'_1(\infty)  \, .
  \end{split}
\end{align}
Here, $\zeta_R^\prime \left(s, a \right)$ is the derivative over $s$ of the generalized Riemann zeta function and
 $\gamma_E$ is the Euler's constant.
The three roots $a_i$ of the polynomial in~\eqref{eqRlinf} are
\begin{equation} \label{eqaChi}
  a_i = 1 - \left(c_2 + \chi_i \left(c_2^2 - 3 c_1 \right)/\theta + \chi_i^* \theta \right)/3 \, ,
\end{equation}
with $\theta^3 = 9/2\left( c_1 c_2 - 3 c_0 - 2/9 c_2^3 + \sqrt{(27 c_0^2 + 4 c_1^3 -18 c_0 c_1 c_2 - 
c_1^2 c_2^2 + 4 c_0 c_2^3)/3}  \right)$ and $\chi = \{ -1, (1\pm i \sqrt 3)/2 \}$, while
$c_i$ are given in~\eqref{eqc0}-\eqref{eqc2}.
This completes the finite part of the decay rate.
Next, we are going to recover the asymptotic terms $\mathcal R_l^a$ that were subtracted 
in~\eqref{eq:zetafiniteRl} using the zeta function regularization.

%
% Zeta function regularization
%
\section{Zeta function regularization} \label{sec:ZetaRenorm}
The decay rate in~\eqref{eqGammaFV} is a physical quantity that depends on the parameters
of the potential $V(\varphi)$ in $S[\varphi]$.
These need to be renormalized to make connections between measurements, such as decay 
rates and scattering cross sections, observed at the minimum of the potential.
Most commonly, the renormalization is done perturbatively via Feynman diagrams and dimensional 
regularization. 
It introduces an arbitrary renormalization scale $\mu$ to keep the mass dimensions 
for any $D$ and ascribes $1/(4-D)$ poles to divergent parts of the momentum integrals.
Within a chosen subtraction scheme, such as $\overline{\text{MS}}$, on-shell or other, 
the renormalized parameters (or counterterms) will remove infinities in physical quantities.

The above holds for the FV decay rate in~\eqref{eqGammaFV} as well~\cite{Coleman:1978ae}. 
One can compute the UV part of the determinant with Feynman diagrams~\cite{Baacke:2003uw, 
Dunne:2005rt, Endo:2017gal, Endo:2017tsz, Oda:2019njo, Andreassen:2017rzq} for scalars, fermions and gauge bosons.
The counterterms used for other processes, will also make the effective action and 
therefore the rate, finite.
For the effective action, which describes the UV part of the FV decay rate, to be consistent
with the finite sum over the eigenvalues, the asymptotic parts are computed by expanding in 
terms of $V^{\prime \prime}(\rho)$ insertions.
In the SM this is equivalent to insertions of the quartic and gauge couplings, which defines
the power counting.

% Connect to Dunne, Min and all the zeta
Alternatively, the UV part of the determinant can be defined by the zeta function~\cite{Minakshisundaram:1949xg, 
Hawking:1976ja}, see~\cite{Elizalde:1994gf} for a review\footnote{For a pedagogical introduction with 
examples regarding the use of spectral functions/functional determinants in physical settings, 
see~\cite{Kirsten:2000ad, Kirsten:2001wz, Kleinert:2004ev, Dunne:2007rt}.}. 
The zeta function formalism was applied to FV decay in the early works~\cite{Konoplich:1987yd}
and more recently in~\cite{Dunne:2006ct}.
We will review its introduction via the contour integral~\cite{Kirsten:2000ad, Kirsten:2001wz, Kirsten:2004qv,
Kirsten:2003py, Kirsten:2005di} in the following section.
Similarly to dimensional regularization, the renormalization scale is introduced for dimensional reasons
to define the zeta function for any value of its argument.
As with Feynman diagrams, the UV part is computed perturbatively in powers of $V^{\prime \prime}$.
However, contrary to the diagrammatic approach, the UV part of zeta is an expansion in powers of 
$l$ and therefore serves as a convenient UV regulator.
Finally, the renormalization is performed by an analytic continuation of the zeta function
and follows from the analyticity of the Riemann zeta function.
We will see that the final result for the single quartic rate via Feynman diagrams agrees with the
zeta function approach.

%
% Zeta function formalism
%
 \subsection{Zeta function via contour integral} \label{sec:ZetaContour}
Let us begin by redefining the sum over the eigenvalues of $\mathcal O$ in terms of the zeta function
\begin{align}\label{eq:detOzeta}
  \ln \det {\mathcal O} = \sum_{\bf n} \ln \gamma_{\bf n} = & -\frac{\text{d}}{\text{d} s}  \sum_{\bf n} \left(  
  \frac{\mu^2}{\gamma_{\bf n}}\right)^s \Bigr \vert_{s=0} 
= -\frac{\text{d}}{\text{d} s} \left( \mu^{2s} \, \zeta_{\mathcal O} (s) \right) \Bigr \vert_{s=0} \, ,
\end{align}
where ${\bf n}$ stands for all the quantum numbers and $\mu$ is the renormalization scale, which keeps the sum
over eigenvalues dimensionless for all values of $s$. 
As found in~\cite{Dunne:2006ct}, it corresponds to the same scale arising from dimensional 
regularization in the $\overline{\text{MS}}$ scheme~\cite{Baacke:2003uw}. 
The zeta function associated to the ratio of determinants is given by the difference
\begin{align}\label{eq:detZeta}
  \zeta &=  \mu^{2s}  \left(  \zeta_{\mathcal O} -\zeta_{\mathcal O_{\text{FV}}} \right) 
  & \text{ and } & &
  \ln\left(\frac{  \det \mathcal O}{\det \mathcal O_{\text{FV}}} \right)  &= -\frac{\text{d}}{\text{d} s} \zeta(s) \Bigr \vert_{s=0} \, .
\end{align}
The sum over eigenvalues in~\eqref{eq:detOzeta} converges if $\operatorname{Re}(s)> D/2$~\cite{Weyl:1912}.
However, to analytically continue $\zeta$ to the region of interest $s=0$, we have to regularize the integral.

\begin{figure}[!ht]
  \centering
  \includegraphics[width=.5 \columnwidth]{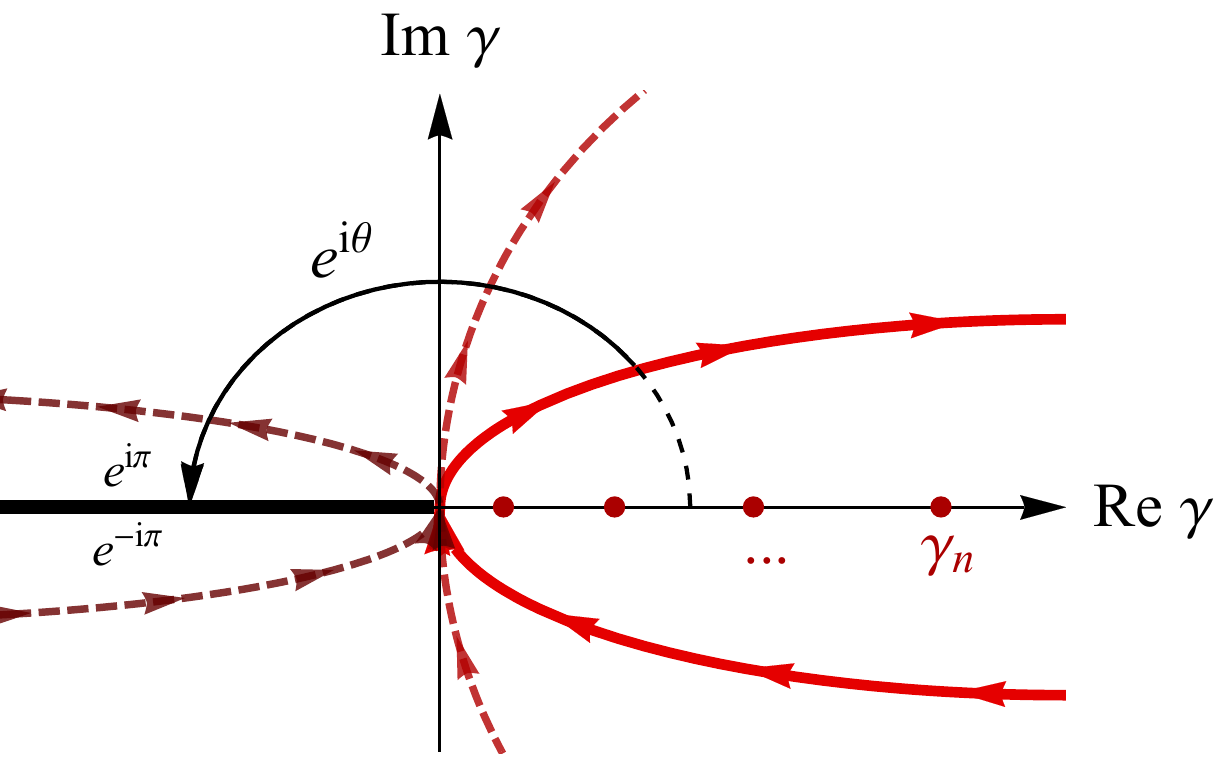}%
  \caption{
     Deformation of the integration contour in~\eqref{eqZetaGen} from the positive real axis to the negative one. 
     The red dots represent the location of the poles, such that $\psi(\infty, \gamma) = 0$. }
  \label{fig:residualTheorem}
\end{figure}

To obtain the analytical structure of $\zeta$ in the range $\operatorname{Re}(s) \leq 2$, we  
rewrite the sum in~\eqref{eq:detOzeta} as a contour integral. 
For this purpose, let us consider $\mathcal O \psi(\rho, \gamma) = \gamma \psi(\rho, \gamma)$, where 
$\gamma$ is a continuous complex parameter.
The $\psi(\gamma)$ is a generalization of $\psi_{\bf n}$ in the sense that when the boundary conditions 
in~\eqref{eq:eigenvalueEquation} are imposed, $\gamma$ becomes quantized and
$\psi_{\bf n}$ is recovered with $\gamma \to \gamma_{\bf n}$.
Now the zeta function is defined as a contour integral
\begin{align} \label{eqZetaGen}
  \zeta_{\mathcal O} &= \sum_{\bf n} \frac{1}{\gamma_{\bf n}^s}
  = \frac{1}{2 \pi i} \oint \frac{\text{d} \gamma}{\gamma^s} \frac{\text{d}}{\text{d} \gamma} \ln \psi (\infty, \gamma)  \, .
\end{align}
The sum over eigenvalues $\gamma_{\bf n}^{-s}$ is recovered because the simple poles are set by 
$\text{d}\ln \psi/\text{d} \gamma = \psi^\prime/\psi$ and the boundary condition
$\psi(\infty, \gamma) \xrightarrow{\gamma \to \gamma_{\bf n}} 0$.
Thus, by the residue theorem, the integral in~\eqref{eqZetaGen} sums up all the eigenvalues, as long 
as the integration contour runs counterclockwise and encloses the entire real axis, as shown by the solid 
red line in Fig.~\ref{fig:residualTheorem}.

As explained above, we have to deform the contour from the positive real axis, which 
encloses all the eigenvalues, to the negative one.
For this purpose, we split the contour in two paths, parametrized in the complex plane by $\exp(\pm i \theta) \gamma$. 
As shown in Fig.~\ref{fig:residualTheorem}, we start with a path that runs along the positive real axis,
\begin{align}
  \zeta_{\mathcal O} &= \frac{1}{2 \pi i} \left(
     \int_0^{\infty} \text{d} \gamma \frac{e^{i \theta s}}{\gamma^s} \frac{\text{d}}{\text{d} \gamma} \ln \psi (\infty, e^{-i \theta} \gamma)  +
     \int_{\infty}^0 \text{d} \gamma \frac{e^{-i \theta s}}{\gamma^s} \frac{\text{d}}{\text{d} \gamma} \ln \psi (\infty, e^{i \theta} \gamma) \right) ,
\end{align}
and deform it to the negative real axis by taking the limit $\theta \to \pi$. We get
\begin{align}\label{eq:zetaO}
   \zeta_{\mathcal O} = \frac{\sin \pi s}{\pi} \int_0^\infty \frac{\text{d} \gamma}{\gamma^s} \frac{\text{d}}{\text{d} \gamma} \ln \psi (\infty, - \gamma) \, ,
\end{align}
where we assume that $\psi$ is continuous around the negative real axis, such that 
$\psi(\rho, e^{\pm i \pi} \gamma) = \psi(\rho, - \gamma)$.
Finally, since we are considering a hyperspherically symmetric potential, we can separate the variables 
\begin{equation} \label{eqOpGammal}
  \mathcal O_l \psi_l(\rho, \gamma) = \gamma\, \psi_l (\rho, \gamma) \, ,
\end{equation} 
and take into account the degeneracy of the orbital modes. Using~\eqref{eq:detZeta} 
and~\eqref{eq:zetaO}, the zeta function for the ratio of determinants is
\begin{align} \label{eqZeta}
  \zeta &= \frac{\sin \pi s}{\pi} \mu^{2s} \sum_\nu \nu^2 \int_0^\infty \frac{\text{d} \gamma}{\gamma^s} 
    \frac{\text{d}}{\text{d} \gamma} \ln \left( 
    \frac{\psi_l(\infty,-\gamma)}{\psi^{\text{FV}}_l(\infty, -\gamma)}\right) \, .
\end{align}

Alas, a closed form solution of $\psi_l(\rho, \gamma)$ cannot be obtained in general, even for the single 
quartic potential.
Since we are only interested in the asymptotic behavior $\psi_l(\infty, \gamma)$ near the FV, it is enough 
to consider the expansion around the FV, where the solution of~\eqref{eqOpGammal} is
\begin{align} \label{eq:psiFVgamma}
  \psi^{\text{FV}}_l(\rho, -\gamma) &=  I_\nu \left(\sqrt \gamma \rho \right) / \rho \, .
\end{align}
The $K_\nu$ part is discarded due to regularity at $\rho = 0$ and the normalization
factor is chosen to be one. 
When $\rho \to \infty$, the fluctuation potential approaches the FV\footnote{
In general, one should subtract $V^{\prime\prime}_{\text{FV}}$ from $V^{\prime\prime}$ and 
shift the eigenvalues $\gamma \to \gamma - V^{\prime\prime}_{\text{FV}}$ in~\eqref{eq:zetaO} 
and~\eqref{eqOpGammal}, modifying the lower limit of integration. For quartic potentials $V^{\prime\prime}_{\text{FV}} = 0$. }
and we can set up an approximate solution
\begin{align}\label{eq:psigamma}
  \psi_l(\rho, -\gamma) &\simeq f_l \left(\gamma \right) \, \psi^{\text{FV}}_l(\rho, -\gamma) \, ,
\end{align}
where $f_l\left( \gamma \right)$ is a constant to be determined in the section below. 
The $K_\nu$ term was neglected, because it vanishes in the asymptotic limit. 
With this ansatz,~\eqref{eqZeta} becomes
\begin{align} \label{eqZetaContour}
  \zeta &= \frac{\sin \pi s}{\pi} \mu^{2s}  \sum_\nu \nu^2 \int_0^\infty \frac{\text{d} \gamma}{\gamma^s} 
    \frac{\text{d}}{\text{d} \gamma} \ln f_l\left(\gamma\right) \, ,
\end{align}
which is valid for $\operatorname{Re}(s) > 2$.
In order to make it well defined around $s = 0$, we have to find the asymptotic form of
$f_l$ and renormalize it.

%
% Regularization and Finite Contributions
%
\subsection{Renormalization of the functional determinant}\label{sec:ZetaRenorm}

To perform the analytical continuation of $\zeta$ to $s = 0$, we define its asymptotic limit
by expanding~\eqref{eqZetaContour} in the large $l$ limit
\begin{align} \label{eqzetaasy}
  \zeta_a &= \frac{\sin \pi s}{\pi} \mu^{2s} \sum_\nu \nu^2
  \int_0^\infty \frac{\text{d} \gamma}{\gamma^s} \frac{\text{d}}{\text{d} \gamma} \ln f_l^a(\gamma)\, ,
\end{align}
and compute $f_l^a$ perturbatively by expanding around the FV.
Once we have $\zeta_a$, we subtract it from $\zeta$, which removes the leading $l$ divergence 
and produces the finite result
\begin{align}\label{eq:zetatotal}
  \zeta_f = \zeta - \zeta_a \, ,
\end{align}
similarly to what was done for the finite sum in Sec.~\ref{sub:FiniteSum}.
Finally, the divergent terms in $\zeta_a$ will be renormalized using the analytic properties of
the Riemann zeta function. 

%
% Finite Contributions
%
\subsubsection{Asymptotic expansion of the zeta} \label{sec:ZetaAsymp}
As discussed above, we would like to compute~\eqref{eqzetaasy} by considering a double
expansion.
First, $\rho \to \infty$ in~\eqref{eqZeta}, which allows us to construct an implicit iterative solution 
around the FV for a fixed angular mode $l$.
Then the high-$l$ expansion can be performed and we end up with a closed form expression for $\zeta_a$.

{\bf False vacuum expansion.}
When approaching the FV,~\eqref{eqOpGammal} can be solved by starting from 
$\psi_l^\text{FV} (\rho,-\gamma)$, given by~\eqref{eq:psiFVgamma}, and writing down the general ansatz
\begin{align} \label{eq:psilGamAs}
  \psi_l(\rho,-\gamma) &= \psi_l^{\text{FV}} (\rho,-\gamma)+ \int_0^\rho \text{d} \rho_1  
  G(\rho, \rho_1) V^{\prime\prime}(\rho_1) \psi_l(\rho_1, -\gamma) \, ,
  \\
  G(\rho,\rho_1) &= \frac{\rho_1^2}{\rho} \left(I_\nu( \sqrt{\gamma} \rho) K_\nu( \sqrt{\gamma} \rho_1) -
  I_\nu ( \sqrt{\gamma} \rho_1 )K_\nu ( \sqrt{\gamma} \rho)\right) \, ,
\end{align}
where $G$ is the Green function associated with $\mathcal O_l$.
This is a self-referential integral equation, which can be solved iteratively
by starting in the FV and expanding in powers of $V^{\prime \prime}$.
The iteration stops when the zeta function becomes well defined in the asymptotic UV limit and
describes all the high $l$ modes.

Actually, we already know from the normalization in~\eqref{eqLogDetO}, and the discussion
regarding the finite sum in Sec.~\ref{sub:FiniteSum}, that the asymptotic terms need to go up to 
$\nu^{-3}$.
In the doubly asymptotic limit when $\rho, \nu \to \infty$, the Green function is proportional to 
$1/\nu$, which follows from the properties of Bessel functions in the Appendix~\ref{sec:Approximations}.
Thus, each insertion of $V^{\prime\prime}$ in~\eqref{eq:psilGamAs} comes with a factor of $1/\nu$
and it is enough to expand the zeta up to $\mathcal O(V^{\prime\prime 3})$.
Using~\eqref{eq:psigamma} and \eqref{eq:psilGamAs}, we have
\begin{align}  
  \frac{\psi_l(\infty,-\gamma)}{\psi^{\text{FV}}_l(\infty, -\gamma)} =  f_l(\gamma) &= 1 + 
  \int_0^\infty \text{d} \rho \rho^2 K_\nu(\sqrt \gamma \rho) V^{\prime\prime}(\rho) \psi_l(\rho,-\gamma) \, ,
  \\
  &= 1 + f_l^{(1)} + f_l^{(2)} + f_l^{(3)} + \mathcal O (V^{\prime\prime 4}) \, ,
\end{align}
while expanding the log to the same order gives
\begin{equation}\label{eq:lnfl}
   \ln f_l\left(\gamma\right) \simeq f_l^{(1)} - \frac{1}{2} \left( f_l^{(1)2} - 2 f_l^{(2)}\right)
   + \frac{1}{3} \left( f_l^{(1)3} - 3 f_l^{(1)} f_l^{(2)} + 3 f_l^{(3)}\right) \, .
\end{equation}
The integrals $f_l^{(n)}$ are obtained by iterating~\eqref{eq:psilGamAs}
\begin{align} \label{eqIntfl1}
  f_l^{(1)} &= \int_0^\infty \text{d} \rho_1 \rho_1 V^{\prime\prime} (\rho_1) K_\nu(\sqrt \gamma \rho_1) 
  I_\nu( \sqrt \gamma \rho_1) \, ,
  \\ \label{eqIntfl2}
  f_l^{(2)} &=  \int_0^\infty \text{d} \rho_1 \rho_1^2 V^{\prime\prime}(\rho_1) K_\nu(\sqrt \gamma \rho_1) 
  \int_0^{\rho_1} \text{d} \rho_1 G_{12} V^{\prime\prime}(\rho_2) \frac{I_\nu ( \sqrt \gamma \rho_2)}{\rho_2}  \, ,
  \\ \label{eqIntfl3}
  f_l^{(3)} &= \int_0^\infty \text{d} \rho_1 \rho_1^{2} V^{\prime\prime} (\rho_1) K_\nu( \sqrt \gamma \rho_1) 
  \int_0^{\rho_1} \text{d} \rho_1 G_{12} V^{\prime\prime}(\rho_2)
  \int_0^{\rho_2} \text{d} \rho_2 G_{23} V^{\prime\prime}(\rho_3) \frac{I_\nu( \sqrt \gamma \rho_3)}{\rho_3} \, ,
\end{align}
where $G_{ij} = G(\rho_i, \rho_j)$.
This concludes the FV expansion in $V^{\prime \prime}$ and we can focus on isolating the 
high-$l$ behavior.

%
%  High-l expansion
%
{\bf High-l expansion.}
We would like to expand $f_l^{(i)}$ for high $l$ up to $\mathcal O(\nu^{-3})$, while keeping $\rho \to \infty$.
To this end, we evaluate the Bessel functions in~\eqref{eqIntfl1}-\eqref{eqIntfl3}
in the limit when $\nu, \rho \to \infty$ with $\sqrt \gamma \rho/\nu$ fixed,
and use the saddle point approximation, 
see~\eqref{eq:IK}-\eqref{eq:SaddlePoint} in the Appendix~\ref{sec:Approximations} for technical details.

For continuous $V^{\prime \prime}$, the integrals in \eqref{eqIntfl1}-\eqref{eqIntfl2} were calculated 
by~\cite{Dunne:2005rt, Dunne:2006ct} and~\eqref{eqIntfl3} was not needed.
Here, we extend the analysis to take into account the delta function
\begin{align}\label{eq:d2Vbounce}
   V^{\prime\prime} \left( \rho \right) = \sum_s
   V^{\prime\prime}_s(\rho) H \left( (-1)^s ( \rho-R_T) \right) - \mu_V \delta(\rho-R_T) \, .
\end{align}
Performing the integrals~\eqref{eqIntfl1}-\eqref{eqIntfl3} requires some effort and we leave 
the details to the Appendix~\ref{sec:DetailsAsymp}.
The final result up to $\mathcal O \left(\sqrt{\gamma}/\nu\right)^4$ is fairly compact
\begin{align}
\begin{split} \label{eq:lnfla}
  \ln f_l^a &= \sum_s \int_0^\infty \text{d} \rho \rho V_s^{\prime \prime} \left( \frac{t}{2 \nu} + \frac{t^3}{16 \nu^3 } 
  \left( 1 - 6 t^2 + 5 t^4 - 2 \rho^2 V_s^{\prime\prime}  \right) \right) H \left(\left( -1\right)^s \left( \rho-R_T \right) \right)
  \\
  & - \mu_V R_T \left( \frac{t}{2\nu} + \frac{t^3}{16 \nu^3 }\left(1 - 6 t^2 + 5 t^4\right) 
   + \mu_V R_T \frac{t^2}{8\nu^2} \right.
  \\
  &  \left.+ 
  \left( \mu_V R_T\right)^2 \frac{t^3}{24\nu^{3}} \left( 1 -\frac{3}{\mu_V^{2}}  \left(V_1^{\prime \prime} +
  V_2^{\prime \prime} \right) \right) \right) \bigg \vert_{\rho = R_T} \, ,
\end{split}
\end{align}
where $t = (1 + \gamma \left(\rho /\nu\right)^2)^{-1/2}$. 
The first line corresponds to the continuous part of $V^{\prime \prime}$ and reproduces the known results 
of~\cite{Dunne:2006ct} when $V^{\prime \prime}_\text{FV} = 0$.
The terms proportional to $\mu_V$ are new because of the presence of the delta at $R_T$.
This completes the asymptotic description of zeta and~\eqref{eq:lnfla} can be used to evaluate the finite 
sum and carry out the renormalization.

%
% Finite Contributions
%
\subsubsection{Regularization of the finite zeta} \label{sec:ZetaFinite}
The asymptotic form of the zeta function allows us to regulate the large $l$ infinities and
compute the finite sum, similarly to what we did in Sec.~\ref{sub:FiniteSum}.
From~\eqref{eqzetaasy} and \eqref{eq:zetatotal}, we have
\begin{align} \label{eqzetafinite}
  \zeta_f &= \frac{\sin \pi s}{\pi} \sum_\nu \nu^2  \mu^{2s} \int_0^\infty 
  \frac{\text{d} \gamma}{\gamma^s} \frac{\text{d}}{\text{d} \gamma} \left( \ln f_l(\gamma) - \ln f_l^a(\gamma)\right )\, ,
\end{align}
which is finite and analytic in the neighborhood of $s=0$. 
This means we can take the derivative with respect to $s$ and evaluate $\zeta_{f}^{\prime}\left(0\right)$.
In doing that, the terms proportional to $\sin \pi s$ vanish, $\gamma^{-s}$ goes to one
and the integral can be computed trivially by evaluating the terms on the boundaries.

On the upper limit $\gamma \to \infty$ and $V''(\rho)$ in~\eqref{eqOpGammal} vanishes, 
thus $\psi_l(\rho, \gamma)$ goes to the FV solution and $f_l(\gamma \to \infty) \to 1$ for 
both log terms in~\eqref{eqzetafinite}, which go to zero.
This leaves us with the two terms on the lower boundary, when $\gamma \to 0$ (and $\rho \to \infty$, as usual).
First, from the definition of $f_l(\gamma)$ in~\eqref{eq:psigamma} and from~\eqref{eqOpGammal}, it becomes
clear that we end up with the same equation~\eqref{eqDefOl} that defined $\mathcal R_l(\infty)$.
In other words, $f_l(0) = \mathcal R_l(\infty)$.
Second, we need to evaluate the asymptotic part $f_l^a(0)$ by setting $\gamma = 0$ in~\eqref{eq:lnfla}
which sets $t = 1$ and we can integrate over $\rho$ for a specific fluctuation potential.
Now, the finite sum can be performed and we reproduce $\Sigma_f$ in~\eqref{eq:zetafiniteRl}, such that
\begin{align}\label{eq:zetafinitezero}
  -\zeta_f^\prime(0) &= \sum_\nu \nu^2 \left( \ln \mathcal R_l (\infty) - \ln f_l^a(0) \right) = \Sigma_f \, , 
\end{align}
for the single and the quartic-quartic potential.
As a very nontrivial cross-check of the asymptotics, we find that $f_l^a$ computed 
from~\eqref{eq:lnfla}, which is defined directly in terms of $V^{\prime \prime}$, is precisely equal to
the one from $\mathcal R_l(\infty)$ in~\eqref{eqRlab}, i.e. $f_l^a (0) = \mathcal R_l^a(\infty)$.

The procedure that gave~\eqref{eq:zetafinitezero} does not always reproduce the finite sum $\Sigma_f$.
In particular, when $V^{\prime \prime}_\text{FV} \neq 0$, the lower limit of integration over $\gamma$ is shifted
from 0 to $\sqrt{V^{\prime \prime}_\text{FV}}$, and we have to evaluate $f_l^a(\sqrt{V^{\prime \prime}_\text{FV}})$.
In this case, additional terms appear in~\eqref{eq:lnfla} because $t \neq 1$.
However, this is an oversubtraction~\cite{Dunne:2006ct} - such terms are suppressed by $1/\nu^4$ or more 
and get canceled by the renormalized parts below.

%
% Renormalization of the divergences
%
\subsubsection{Renormalization of the asymptotic zeta} \label{secRenorm}
The asymptotic part of the zeta function can now be renormalized. 
The integrals in~\eqref{eq:lnfla} are evaluated using the following identity, valid for $\operatorname{Re}(s)<1$
\begin{align}\label{eq:integral}
 \frac{\sin \pi s}{\pi} \mu^{2s} \int_0^\infty \frac{\text{d} \gamma}{\gamma^s} \frac{\text{d}}{\text{d} \gamma} t^n = 
  -\frac{\Gamma\left( s+\frac{n}{2} \right) \left(\mu \rho\right)^{2s}}{\Gamma\left(s\right) 
   \Gamma\left( \frac{n}{2}\right)} {\nu^{-2s}} \, .
\end{align}
The resulting expressions are plugged into~\eqref{eqzetaasy} and we perform the sum over $\nu$.
Each term that goes as $(t/\nu)^n$ gives a Riemann zeta $\zeta_R(2s + n - 2)$.
The analytic continuation properties of $\zeta_R$ are well known and provide a mathematical description 
of divergencies.
Finally, we take the derivative over $s$ and send $s$ to zero, ending up with
\begin{align}
\begin{split} \label{eqdzetaasy}
  \zeta^\prime_a(0) &= \sum_s \frac{1}{8} \int_0^\infty \text{d} \rho\, \rho^3 V_s^{\prime \prime 2}  
  \left( \ln \left( \frac{\mu \rho}{2}\right) + \gamma_E + 1\right) H \left( \left( -1 \right)^s \left( \rho-R_T \right) \right)
  \\
    & - \frac{\left( \mu_V R_T \right)^2}{16} + \frac{\left( \mu_V R_T \right)^3}{24} \left( 1 - \frac{3}{\mu_V^2}  \left(V_1^{\prime \prime} 
  + V_2^{\prime \prime} \right) \big \vert_{R_T} \right) \left( \ln \left( \frac{\mu R_T}{2}\right) + \gamma_E + 1 \right) \, .
\end{split}
\end{align}
This agrees with~\cite{Dunne:2006ct} for a continuous $V^{\prime \prime}$ with $\mu_V = 0$ and 
$V^{\prime \prime}_\text{FV} = 0$ and also reproduces the SM~\cite{Andreassen:2017rzq}
when applied to the single quartic.
This also demonstrates that the Feynman diagrammatic approach coincides with the zeta function formalism.
Another nontrivial check regards the cancellation of divergences, i.e. we verify that terms proportional to 
$\gamma_E$ in $\Sigma_f$ given by~\eqref{eqSigmafinite} cancel the ones in $\zeta^\prime_a(0)$ above.

Let us comment on the renormalization scale dependence.
The FV decay is a physical process and the rate should not depend on $\mu$.
Specifically, the $\mu$ dependence from the prefactor cancels the $\ln \mu$ from running 
of parameters in the bounce action $\mathcal S_0$.
For the single quartic case this is easy to see. 
The first segment gives $1/2 \zeta_a^\prime(0) \supset 3 \ln \mu$, while the running of 
the quartic $\beta_\lambda = \text{d} \lambda/\text{d} \ln \mu = 9 \lambda^2/(8 \pi^2)$ is solved
for $\lambda(\mu)$ and plugged into the bounce action $-8\pi^2/(3 \lambda(\mu))$ to cancel 
the $\mu$ dependence of the prefactor.
While running the quartic-quartic potential couplings is beyond the scope of this work, we confirm that 
the leading order running of $\lambda_{1,2}$ with the above beta functions cancels the $\mu$ 
dependence of the continuous part of~\eqref{eqdzetaasy} in the weakly coupled limit when $x$ 
and $y$ are small.

%
% Summary of decay rates
%
\section{Summary of decay rates} \label{sec:Summary}

The final result for the renormalized log of the functional determinant is
\begin{equation}
  \ln\left(\frac{  \det \mathcal O}{\det \mathcal O_{\text{FV}}} \right) = - \zeta^\prime(0) = - \zeta_f^\prime(0) - \zeta_a^\prime(0) \, ,
\end{equation}
where $-\zeta_f^\prime(0) = \Sigma_f$ can be found in~\eqref{eqSigmafinite} and $\zeta_a^\prime(0)$ in~\eqref{eqdzetaasy}.
Therefore, the total decay rate per 4D unit volume is
\begin{align} \label{eqGammaFVFinal}
  \frac{\Gamma}{\mathcal V} &= \left( \frac{\mathcal S_0}{2\pi} \right)^{2} e^{- \mathcal S_0 + \frac{1}{2} \zeta'(0)} 
  = v^4 e^{- \mathcal S_0 - \mathcal S_1} \, ,
\end{align}
where the $\mathcal S_0$ comes from~\eqref{eq:Action} and $\zeta'(0)$ is the sum of~\eqref{eqSigmafinite} 
and~\eqref{eqdzetaasy}.
As we will see, having a closed form result is particularly useful to study the behavior of the rate in the TW 
limit as well as for the large scale separation $x \gg 1$, corresponding to a rather flat potential.

%
%  Real quartics
%
\subsection{Real quartic}

\begin{figure}
  \centering
  \includegraphics[width=.8\columnwidth]{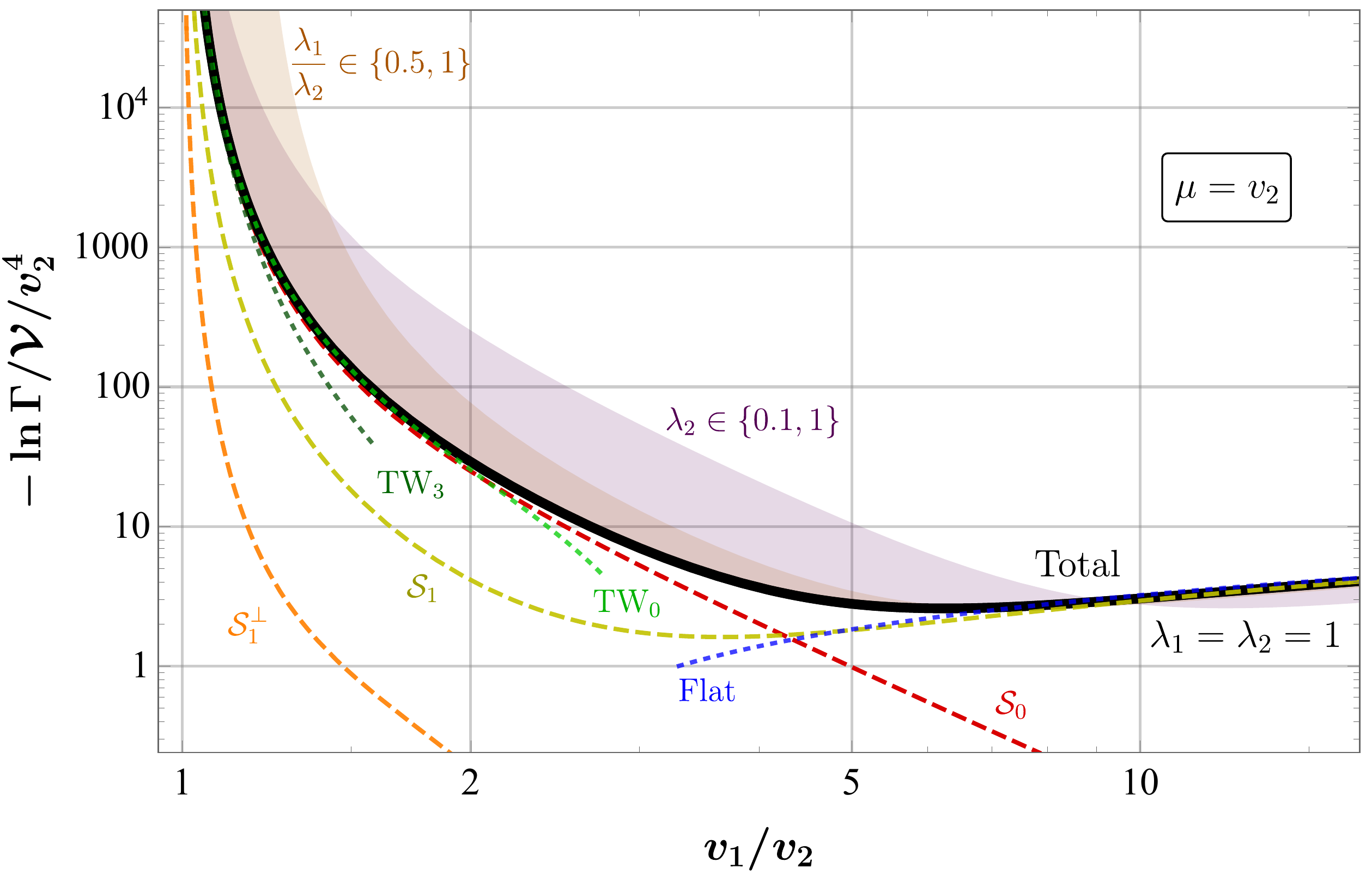}
  \caption{The FV decay rate for the quartic-quartic potential in~\eqref{eqDefV}. 
  The black solid line shows the total rate, while the dashed ones show the semiclassical part $\mathcal S_0$
  in red and the finite renormalized prefactor $\mathcal S_1$ in dark yellow.
  The dotted lines correspond to the TW leading expansion, where we set $y = \lambda_1/\lambda_2 = 1$ and expand up to 
  $(x-1)^{-3}$ in dark green, additional corrections up to $(x-1)^0$ in light green and the flat potential limit $x = v_1/v_2 \gg 1$ in blue. 
  The shaded regions show the variation of $\lambda_2 \in \{0.1 , 1 \}$ in purple and 
  $\lambda_1/\lambda_2 \in \{0.5 , 1 \}$ in light brown.
  }
  \label{fig:GammaFV}
\end{figure}

To complete the calculation for the real scalar part, we evaluate the integral in~\eqref{eqdzetaasy} 
in a closed form with $R_{1,2,T}$ and $\mu_V$ given by~\eqref{eqR12T} and~\eqref{eqMatchRl}, respectively.
This is a straightforward calculation, but we omit the entire expression for brevity\footnote{We provide a 
complete \texttt{Mathematica} notebook with the entire calculation as ancillary material.} 
and instead show the negative
log of the normalized rate $-\ln \Gamma/\mathcal V/v^4$ in Fig.~\ref{fig:GammaFV}.
The total rate is shown by the black solid line on Fig.~\ref{fig:GammaFV} for a fixed $\lambda_1 = \lambda_2 = 1$ as 
$x = v_1/v_2$ interpolates from the TW $x \sim 1$ to the thick wall and a flat potential when $x \gg 1$.
We assume that all the couplings are defined at $v$ and set the renormalization scale to $\mu = v$.
In this case, the rate is insensitive to $v$, apart from the overall normalization factor $v^4$, which is factorized
in the plot.

The contribution from the semiclassical action $\mathcal S_0$ in the first term of~\eqref{eqGammaFVApprox},
coming from the bounce action is shown in dashed red and tends to dominate for small $x$, as long as $\lambda$ is small.
The prefactor correction $\mathcal S_1$ is plotted in dashed yellow.
It is subdominant for small values of $x$ and starts to dominate for $x \sim 4$, the rate drops and 
then rises logarithmically.
Strictly speaking, these two are not separate contributions, i.e. running the parameters in $\mathcal S_0$ 
will exactly cancel the $\mu$ dependent part of $\mathcal S_1$.
Of course, once this is done, one has to include all the corrections of the same order, in particular
the $\log R$ and finite parts - they all contribute with the same power of the coupling constant.
The fact that the running of couplings, or alternatively the $\mu$-dependent part of the prefactor,
becomes important for large $x$, is not surprising.
After all, this is happening in the regime $v_1 > v_2$, where a large separation of scales is present.
This is exactly analogous to any quantum field theoretic calculation, where large logs appear when 
scales are separate and one needs to resum them, even in perturbative theories like quantum electrodynamics.

However, this does not imply a breakdown of the semiclassical approximation - the leading bounce
contribution will always dominate for a small enough $\lambda$.
In any case, lowering $\lambda_2$ results in a higher $\mathcal S_0$ which dominates the 
$\mathcal S_1$ for larger values of $x$, as shown by the purple shaded region.
The variation of $y$, on the other hand, results in a shift of the entire curve to larger $x$, as shown
by the brown shaded region, because the thin wall pole in the rate happens when $x^4 y \simeq 1$.

The behavior of the rate simplifies considerably in these two limits.
Near the thin wall $x^4 y \sim 1$ (TW: $x \sim 1 + \varepsilon, y = 1$), the $a_i$ become large and negative, thus the 
asymptotic expansion of $\zeta^\prime(s,3-a_i)$ in~\eqref{eqZetaP0}-\eqref{eqZetaPm2} can be used.
Conversely, the $a_i$ become nearly constant when $x \gg 1$ (flat) and we have
\begin{equation} \label{eqGammaFVApprox}
  -\ln \frac{\Gamma}{\mathcal V} \frac{1}{v^4} \simeq 
  \begin{cases}
   \frac{1}{\varepsilon^3}\left( \frac{2 \pi^2}{3 \lambda} + \frac{2}{9} +  \frac{\pi}{2 \sqrt 3} - 
    \frac{1}{12} \ln \frac{2\lambda v^2}{\mu^2} \right), & \text{TW} \,,
    \\
    \frac{7}{12} -2 \zeta_{R}^{\prime}\left(-1\right) + \frac{1}{3} \ln \frac{y^2 \lambda^2 v^4 x^6}{32\pi^3 \mu^4} , & \text{Flat} \, .
  \end{cases}
\end{equation}
The leading TW functional dependence goes as $\varepsilon^{-3}$, which is the same as in the TW approximation 
of the displaced quadratic potential~\cite{Konoplich:1987yd}, with different numerical coefficients and 
an additional log term.
The TW series can easily be extended to arbitrary order in $\varepsilon$; we plot the leading 
$\varepsilon^{-3}$ and the expansion up to $\varepsilon^0$ with the dotted green lines in Fig.~\ref{fig:GammaFV}.
These two are fairly good proxies and cover a significant portion of parameter space,
as seen from Fig.~\ref{fig:GammaFV}.

%
%  Complex quartics
%
\subsection{Complexified quartic} \label{subsec:CplxQ}
Let us extend the analysis to the complexified version of the model and examine the effect of 
transverse fluctuations, coming from the imaginary field component.

As we will see, when restricted to the single quartic, we are dealing with a global $U(1)$ 
symmetric theory with a massless Goldstone and obtain an additional zero eigenvalue.
This situation is somewhat similar to the SM, however one should be careful with the 
comparison, since the equations of motion and the would-be-Goldstone masses are 
gauge-dependent~\cite{Andreassen:2017rzq, Endo:2017tsz, Ai:2020sru}.
Once we add the second quartic segment, the global $U(1)$ disappears together with the zero 
eigenvalue.

Consider a complex scalar field $\Phi = (\varphi + i \, \varphi_\perp)/\sqrt 2$ and the complexified version 
of the potential
\begin{align} \label{eqDefVCplx}
  V\left( \Phi \right) &= \left( \lambda_2 v_2^4 - \lambda_1 v_1^4 + 
  \lambda_1 \left| \Phi + v_1 \right|^4 \right) H \left(\tilde \Phi - \Phi \right) +
  \lambda_2 \left| \Phi - v_2 \right|^4 H \left(\Phi - \tilde \Phi \right) \, ,
\end{align}
where $\tilde \Phi(\Phi)$ describes the boundary between the two regions.
It is chosen such that $V$ is continuous in the $\varphi-\varphi_\perp$ plane and $\tilde \Phi$
goes to zero on the $\varphi_\perp = 0$ axis, reproducing~\eqref{eqDefV}.
Parameters of the potential are still real and the bounce for the real component 
$\varphi$ stays the same, as does the determinant.

The perpendicular component $\varphi_\perp$ carries no vev, because $v_{1,2} \in \mathbb R$, and
its bounce is zero.
The fluctuations $\psi_l^{\perp}$ are nonzero and obey
\begin{align} \label{eqDefOPerp}
   \mathcal O_\perp \psi_l^{\perp} = -\ddot \psi_l^{\perp} - \frac{3}{\rho} \dot \psi_l^{\perp} + 
   \frac{l(l+2)}{\rho^2} \psi_l^{\perp} + 
   V^{\prime \prime}_\perp \left( \overline \varphi \right) \psi_l^{\perp} &= 0 \, ,
  &
  V^{\prime \prime}_\perp &= \frac{1}{3} V^{\prime \prime} \, .
\end{align}
The FV normalization stays the same $\psi_{l s}^{\text{FV} \perp} = \rho^l$, 
while the transverse fluctuations are simpler than the real scalar ones
\begin{align} \label{eqPsiPerpl}
   \psi_{l s}^\perp & = \frac{\rho^l \, R_s^2}{R_s^2 - \rho^2} 
   \left( A^\perp_{l s} \left( 1 - \left( \frac{l}{l + 2} \right) \frac{\rho^2}{R_s^2} \right) + 
   B^\perp_{l s} \, \frac{R_s^{2 l + 2}}{\rho^{2 l + 2}} \left( 1 - \left(\frac{l+2}{l} \right) \frac{\rho^2}{R_s^2} \right) \right) \, .
\end{align}
The boundary conditions fix $A^\perp_{l1} = 1, B^\perp_{l1} = 0$, such that dividing by $\rho^l$ and taking the 
limit $\rho \to \infty$, we recover the single quartic global Goldstone~\cite{Andreassen:2017rzq}
\begin{equation} \label{eqRlInfqPerp}
  \lambda \left| \Phi \right|^4 : \mathcal R_l^{\perp} (\infty) = \frac{l}{l+2} \, ,
\end{equation}
where the zero eigenvalue at $l=0$ appears due to the $U(1)$ symmetry.
Proceeding to the second segment and taking into account the matching conditions, we end up with
\begin{align} \label{eqRlinfPerp}
  \mathcal R_l^\perp(\infty) &= A^\perp_{l2} \, \frac{l}{l+2} = \frac{l^3 + c^\perp_2 l^2  + c^\perp_1 l+ c^\perp_0}{(l+1) (l+2)^2} \, .
\end{align}
After adding the second segment, the $U(1)$ symmetry gets broken and the zero eigenvalue in~\eqref{eqRlInfqPerp}
disappears. The coefficients are then given by $c^\perp_0 = c_0/9$ and   
\begin{align}
  c^\perp_1 &= \frac{2 x \left((2 x+1) x^2 y+1\right) \left((4 x+1) x^3 y+x-2\right)}{3 \left(x^4 y-1\right)^2} \, , 
  &
  c^\perp_2 &= \frac{(13 x+4) x^3 y+4 x-5}{3 (x^4 y - 1)} \, .
\end{align}
The $\mathcal R_l^\perp(\infty)$ goes to 1 as $l \gg 1$ and the ratio of determinants diverges.
To get the total rate, we proceed as for the real quartic above.
The solutions to the cubic polynomial in~\eqref{eqRlinfPerp} are given by the same expression in~\eqref{eqaChi}
with replacing $c_i \to c_i^\perp$ and the fluctuation potential $V^{\prime \prime}_\perp \to V^{\prime \prime}/3$.
Again, the asymptotic behavior is simple
\begin{equation} \label{eqGammaFVApproxCplx}
  \mathcal S_1^\perp\simeq 
  \begin{cases}
 \frac{1}{972\varepsilon^3}  \left( 152-8\ln 27- 12\sqrt{11} \arctan \sqrt{11} - 
  57 \ln \frac{2\lambda v^2}{\mu^2}  \right) , & \text{TW} \,,
  \\
    \frac{322}{81} \ln 2
    -\frac{1}{324}
    -\frac{\zeta_{R}\left(3\right)}{8\pi^2}
    -\frac{5}{2} \ln 3
    + \frac{2}{81} \ln \frac{\lambda y x^3 v^2}{\mu^2} 
   -\frac{\zeta_{R}^{\prime}\left(-2,\frac{7}{3}\right) }{2} 
   -\frac{2 \zeta_{R}^{\prime}\left(-1,\frac{7}{3}\right) }{3}
   -\frac{2\zeta_{R}^{\prime}\left(0,\frac{7}{3}\right) }{9}
   , & \text{Flat} \, ,
  \end{cases}
\end{equation}
and the total rate is obtained by adding $\mathcal S_1^\perp$ to $\mathcal S_1$ in~\eqref{eqGammaFVFinal}.
It turns out that the correction from the transverse fluctuations are rather small and 
subdominant with respect to the real scalar ones, as seen from the orange dashed
line on Fig.~\ref{fig:GammaFV}.

%
% Conclusions and outlook
%
\section{Conclusions and outlook} \label{sec:Outlook}
%
% ===== Conclusions & Outlook =====
%
We presented a closed-form solution for the total decay rate at one loop for a potential with 
two tree level minima of a quartic-quartic potential.
Our approach is based on the Gel'fand-Yaglom theorem that circumvents the need to obtain
individual eigenvalues of the fluctuation operator.
The existing renormalization procedure had to be generalized 
to include the delta functions in the fluctuation potential.
To this end, an appropriate expansion of the fluctuation functions to the maximal $1/l^3$ 
term had to be performed to extract the UV behavior and regularize the determinant.
%
% Extension to other renormalization procedures that deal with delta functions.
% Feynman diagram approach and the RGEs
%
It might be of interest to reproduce this result with the Feynman diagrammatic approach
and also obtain the RGE running of parameters for this particular case.

The final expression for the FV decay rate in~\eqref{eqGammaFVFinal} consists of the semiclassical
action $\mathcal S_0$ and the finite and renormalized corrections $\mathcal S_1$.
Both are calculated for a complete range of parameters of the potential -- thus we get an exact one loop
result for the renormalized decay rate.
The main result of our work is summarized in Fig.~\ref{fig:GammaFV}, where the behavior of the rate for thin 
and thick walls becomes apparent, as well as the range of validity of the simple approximations that were 
derived from the exact result.
Perhaps the main take-away message here is that for a large separation of scales, one has to include
the running of parameters in $\mathcal S_0$ and subsequently compute the renormalization scale dependent
prefactor, such that the total rate becomes $\mu$-independent.
Note that this does not signal the breakdown of perturbativity, which instead is governed by the overall
size of the quartic coupling $\lambda$.

We also included the effects of the imaginary component of the complex field, which are found to
be subdominant in general.
Similarly to the SM, which corresponds to the single quartic, the effect of fermions and gauge
bosons could be taken into account.
To this end, the known results~\cite{Endo:2017gal, Endo:2017tsz, Andreassen:2017rzq}, for spin 1/2 and 
1 fluctuations should be extended to include the second quartic segment while taking into account the 
presence of Dirac delta, in complete analogy to the imaginary complex scalar.

% Generalization to an arbitrary number of dimensions.
%
The present calculation relies on an exact bounce solution that can be found in $D=4$.
It may be of interest to extend its validity via dimensional continuation to other dimensions, $D=3$ in particular.
This may be possible to do perturbatively, similarly to the bounce~\cite{Konstandin:2006nd}, near the thin wall, 
where the $1/\rho$ term does not play a significant role.
Likewise, one may consider other solvable bounces, such
as the log potential~\cite{Shen:1988si,FerrazdeCamargo:1982sk}, 
quadratic-quadratic~\cite{Pastras:2011zr}, binomial~\cite{Aravind:2014pva} and 
(extended) polygonal~\cite{Guada:2018jek}.
The latter is particularly interesting because the fluctuation potential is smooth and avoids the delta function.
At the same time, it can serve as a universal estimator of the total rate and might be extended to
multifields~\cite{Guada:2020wip}, where only recently~\cite{Chigusa:2020pr} progress was made.

% Exact solutions of the fluctuation operator: 
%
% - Log Potential in D = 3 and 4, 
% - Quadratic - Quadratic in D = 3 and 4,
% - Binomial Potential in D = 4, 
% - Extended Polygonal
% - Nontrivial potentials, disappearing bounces, intermediate minima
%
% Other transverse fluctuations: fermions and gauge bosons
% Non trivial bounce solution in field space: Multi Field Potentials (work in progress)

% There are no tools for the fluctuations on the market
% 

\acknowledgments 
We thank Rok Medve\v s for discussions and work on topics related to this work.
The work of V.G. was supported by the Slovenian Research Agency's young researcher program 
under the grant No. PR-07582. 
M.N. was supported by the Slovenian Research Agency under the research core funding No. P1-0035 and 
in part by the research grant J1-8137.
M.N. acknowledges the support of the COST action CA16201 - ``Unraveling new physics at the LHC 
through the precision frontier''.
  
%
% Appendices
%
\appendix
\renewcommand*{\thesection}{\Alph{section}}
\section{Bessel, Saddle-point and Zeta function approximations} \label{sec:Approximations}
%
% Bessel functions
%
{ \bf 1. Bessel functions.}
To perform the high-$l$ expansion in Sec.~\ref{sec:ZetaAsymp}, we used the mathematical properties of the Bessel 
functions that can be found on p.~378 of~\cite{Abramowitz:1974} Eqs.~(9.7.7) and (9.7.8). 
Expanding for large $\nu$ and $\rho$, while keeping $\rho/\nu$ fixed, we have up to $\mathcal O\left(\nu^{-4}\right)$
\begin{align} \label{eq:IK}
  I_\nu(\sqrt \gamma \rho) K_\nu(\sqrt \gamma  \rho) &= \frac{t}{2\nu} + \frac{t^3}{16 \nu^3}\left(1-6 t^2 +5 t^4\right) \, ,
\end{align}
and up to $(1 + \mathcal O \left(\nu^{-1} \right))$
\begin{align}
  \label{eq:I2K2prime}
  I_\nu^2 \left(\sqrt \gamma \rho \right) &= \frac{t}{2\pi\nu}{e^{2 \nu \eta }} \, ,
  \quad
  K_\nu^2 \left(\sqrt \gamma \rho \right) \sim \frac{\pi\,t}{2\nu}{e^{-2 \nu \eta}} \, ,
\end{align}
with $\eta = t^{-1} + \ln \left(  \sqrt \gamma \rho/\nu/ \left(1+t^{-1}\right) \right)$.

%
% Saddlepoint approximation
%
{ \em Saddle-point approximation}
can be found on p. 362 of~\cite{Kirsten:2000ad,Kirsten:2001wz} Eq.~E.14. It can be used to expand the integrals 
in Sec.~\ref{sec:ZetaAsymp} in powers of $1/\nu$ when the leading contribution 
is dominated by the exponential high-$l$ terms from~\eqref{eq:I2K2prime}. Expanding up to $ \mathcal O \left(\nu^{-2} \right)$
\begin{align}\label{eq:SaddlePoint}
   \int_0^\rho \text{d} \rho_1 f(\rho_1) e^{\nu B(\rho_1)} =  e^{\nu B(\rho)}   
   \frac{f(\rho)}{\nu} \left( \frac{\text{d} B(\rho)}{\text{d} \rho} \right)^{-1}\, .
\end{align}

%
% Generalized Riemann Zeta function
%
{\bf 2. Generalized Riemann Zeta function}
A useful asymptotic expansion of the derivatives of the generalized zeta function, 
is applicable in the TW limit $a \gg 1$ and can be found in Eqs.~(18) and~(19) of~\cite{Elizalde:1986az}
\begin{align} \label{eqZetaP0}
  \zeta_R^\prime(0, a) &= \ln \Gamma(a) - \frac{\ln 2 \pi}{2} \sim -a + a \log a - \frac{\log a}{2} ,
  \\ \label{eqZetaPm1}
  \zeta_R^\prime(-1, a) &\sim -\frac{a^2}{4}+\frac{a^2 \log a}{2} - \frac{a \log a}{2}  + 
  \frac{\log a}{12} + \frac{1}{12} - \sum_{k=1}^\infty \frac{B_{2 k +2} a^{-2k} }{(2k+2)(2k+1)2k},
  \\ \label{eqZetaPm2}
  \zeta_R^\prime(-2, a) &\sim -\frac{a^3}{9}+\frac{a^3 \log a}{3} - \frac{a^2 \log a}{2} + \frac{a}{12} +
  \frac{a \log a}{6} + \sum_{k=1}^\infty \frac{2 B_{2 k +2} a^{-(2k-1)} }{(2k+2)(2k+1)2k(2k-1)} \,,
\end{align}
where $B_k$ are the Bernoulli numbers.

%
% Derivation of the asymptotic behavior
% 
\section{Derivation of the high-$l$ expansion of $f_l$} \label{sec:DetailsAsymp}
This section is devoted to the derivation of $\ln f_l^a$ in~\eqref{eq:lnfla} from 
the high-$l$ expansion of $\ln f_{l}$ in~\eqref{eq:lnfl} up to $\mathcal O\left(\nu^{-4} \right)$, 
while keeping $\rho \to \infty$.
For this purpose, let us first plug $V^{\prime\prime}$ from~\eqref{eq:d2Vbounce} 
into~\eqref{eq:lnfl} and separate the integrals in three parts: the terms proportional 
to the delta function, to the Heaviside unit step function and the cross terms. 

{\bf Delta function} terms come purely from the discontinuity of the first derivative of 
the potential $\bar\varphi\left( R_T \right) = 0$. One can compute the
integrals exactly and perform the high-$l$ expansion from~\eqref{eq:IK}.
This gives the terms proportional to $\mu_V R_T$ in~\eqref{eq:lnfla}, one 
for each insertion of $V^{\prime\prime}$.
For instance, the last three terms of~\eqref{eq:lnfl}, which are of third order in $V^{\prime\prime}$ are
\begin{align}
   \ln f_l^a &\supset -\frac{1}{3} \left( \mu_V R_T I_\nu(\sqrt \gamma \rho) K_\nu(\sqrt \gamma \rho) \right)^3
   \sim -\frac{1}{3} \left( \frac{t}{2 \nu} \mu_V R_T \right)^3 \, ,
\end{align}
where we kept all the terms up to $\mathcal O \left( \nu^{-4} \right)$.

{\bf Heaviside} terms belong to the continuous part of $V^{\prime\prime}$. 
They were first computed by~\cite{Dunne:2005rt, Dunne:2006ct} and contribute to the first term of~\eqref{eq:lnfla}.
Let us proceed to compute each term of~\eqref{eq:lnfl} by neglecting the delta terms.

The leading order terms in $V^{\prime\prime}$ can be computed simply by using the Bessel 
expansions in~\eqref{eq:IK}.
The second order terms in $V^{\prime\prime}$ can first be simplified by 
\begin{align}
   \int_0^\infty \text{d}\rho_1 \int_{\rho_1}^\infty  \text{d}\rho & = \int_0^\infty \text{d}\rho \int_0^\rho \text{d}\rho_1 \, ,
\end{align}
since $V^{\prime\prime}$ is continuous, as shown in the Appendix E of~\cite{Kirsten:2000ad,Kirsten:2001wz}. 
At $\mathcal O\left( \nu^{-4} \right)$, this leads to
\begin{align}
  \ln f_l^a & \supset \sum_s \int_0^\infty \text{d} \rho \, \rho \int_0^\rho \text{d} \rho_1 \rho_1 
  K^2_\nu(\sqrt \gamma \rho) V^{\prime\prime}_s \left( \rho \right) V^{\prime\prime}_s 
  \left( \rho_1 \right) I^2_\nu(\sqrt \gamma  \rho_1)
  \\
  & \sim \frac{t^3}{8 \nu^3}  \sum_s \int_0^\infty \text{d} \rho \, \rho^3 
  V^{\prime \prime 2}_s H\left( \left(-1 \right)^s \left( \rho - R_T \right) \right) \,,
\end{align}
where we used the exponential behavior of $I_{\nu}$ and $K_{\nu}$ in~\eqref{eq:I2K2prime} and 
the saddle-point approximation~\eqref{eq:SaddlePoint} in the last step.
Finally, the third order terms go as $\mathcal O\left(\nu^{-4} \right)$ and do not contribute 
to $f_l^a$ since each pair of Bessel functions~\eqref{eq:I2K2prime} as well as the saddle-point 
approximation~\eqref{eq:IK} come with a factor of $1/\nu$ .

{\bf Cross terms} require a careful treatment in the integration of the delta function 
since it brings a Heaviside that affects the limits of integration of the second integration.
Then we perform the asymptotic expansion of the Bessel functions and the saddle-point approximation 
as in the previous calculations with~\eqref{eq:I2K2prime} and~\eqref{eq:SaddlePoint}.
These correspond to the last two terms of~\eqref{eq:lnfla}.
For example, the last term of~\eqref{eq:lnfla} is given by 
\begin{align}
\begin{split}
    \ln f_l^a &\supset \sum_s \int_0^\infty \text{d} \rho \rho K_{\nu}^2\left(\sqrt \gamma \rho \right) 
    V^{\prime\prime}_s \left(\rho\right)
   \int_0^\rho d\rho_1 \,\rho_1 \mu_V \delta \left(\rho - R_T \right) I_{\nu}^2\left(\sqrt \gamma \rho_1 \right)  
   \\
   &=  \mu_V R_T I_{\nu}^2\left(\sqrt \gamma R_T \right) \sum_s \int_{R_T}^\infty
   \text{d} \rho \rho K_{\nu}^2 \left(\sqrt \gamma \rho \right) V^{\prime\prime}_s \left( \rho \right) H(\rho -R_T)
   \\
   &\sim  \frac{1}{\mu_V^2} \left(\frac{t}{2 \nu} \mu_V R_T \right)^3 V^{\prime\prime}_2 \left(R_T\right) \, ,
\end{split}
\end{align}
where in the second line, the integration limits have changed due to the previous integration of the delta function,
which picks $V_{2}^{\prime\prime}$ \footnote{This actually depends on our convention of the Heaviside. 
We have chosen that $H(x) = 0$ when $x<0$ but equivalently, we could have used 
$H(x) = 1/2$ when $x=0$ and, after adding up all the cross terms in $\ln f_l^a$, we recover the same results.}
. Then we used the saddle point approximation that 
evaluates the potential at $R_T$ and provides the last line.

The next to last term of~\eqref{eq:lnfla} can be computed completely analogously, while
the remaining terms in~\eqref{eq:lnfl} cancel among themselves or go as $\mathcal O \left( 1/\nu^4 \right)$. 
After collecting all the results, we are left with the final expression given in~\eqref{eq:lnfla}.

\def\arxiv#1[#2]{\href{http://arxiv.org/abs/#1}{[#2]}}
\def\Arxiv#1[#2]{\href{http://arxiv.org/abs/#1}{#2}}

\end{document}